\documentclass[twocolumn,english,reprint, longbibliography, superscriptaddress, breaklinks=true, showkeys, showpacs=false, nofootinbib]{revtex4}
\usepackage[T1]{fontenc}
\usepackage[latin9]{inputenc}
\usepackage{color}
\usepackage{babel}
\usepackage{amsmath}
\usepackage{amssymb}
\usepackage{graphicx}
\usepackage{physics}
\usepackage{subfigure}
\usepackage[unicode=true,pdfusetitle,
 bookmarks=true,bookmarksnumbered=false,bookmarksopen=false,
 breaklinks=true,pdfborder={0 0 0},backref=false,colorlinks=true]
 {hyperref} 
\usepackage{breakurl}
\usepackage{qcircuit}

\makeatletter
\@ifundefined{textcolor}{}
{%
 \definecolor{BLACK}{gray}{0}
 \definecolor{WHITE}{gray}{1}
 \definecolor{RED}{rgb}{1,0,0}
 \definecolor{GREEN}{rgb}{0,1,0}
 \definecolor{BLUE}{rgb}{0,0,1}
 \definecolor{CYAN}{cmyk}{1,0,0,0}
 \definecolor{MAGENTA}{cmyk}{0,1,0,0}
 \definecolor{YELLOW}{cmyk}{0,0,1,0}
}

\pdfoutput=1
\hypersetup{colorlinks=true,citecolor=blue,linkcolor=cyan,urlcolor=blue,filecolor= green, breaklinks=true}
\usepackage{url}
\usepackage{breakurl}

\makeatother

\begin{document}

\title{Experimental tests of density matrix's properties-based complementarity relations}

\author{Mauro B. Pozzobom}
\email{mbpozzobom@gmail.com}
\address{Departamento de F\'isica, Centro de Ci\^encias Naturais e Exatas, Universidade Federal de Santa Maria, Avenida Roraima 1000, Santa Maria, RS, 97105-900, Brazil}

\author{Marcos L. W. Basso}
\email{marcoslwbasso@mail.ufsm.br}
\address{Departamento de F\'isica, Centro de Ci\^encias Naturais e Exatas, Universidade Federal de Santa Maria, Avenida Roraima 1000, Santa Maria, RS, 97105-900, Brazil}

\author{Jonas Maziero}
\email{jonas.maziero@ufsm.br}
\address{Departamento de F\'isica, Centro de Ci\^encias Naturais e Exatas, Universidade Federal de Santa Maria, Avenida Roraima 1000, Santa Maria, RS, 97105-900, Brazil}

\selectlanguage{english}%

\begin{abstract}
Bohr's complementarity principle is of fundamental historic and conceptual importance for Quantum Mechanics (QM), and states that, with a given experimental apparatus configuration, one can observe either the wave-like or the particle-like character of a quantum system, but not both. However, it was eventually realized that these dual behaviors can both manifest partially in the same experimental setup, and, using ad hoc proposed measures for the wave and particle aspects of the quanton, complementarity relations were proposed limiting how strong these manifestations can be.
Recently, a formalism was developed and quantifiers for the particleness and waveness of a quantum system were derived from the mathematical structure of QM entailed in the density matrix's basic properties ($\rho\ge 0$, $\mathrm{Tr}\rho=1$). In this article, using IBM Quantum Experience quantum computers, we perform experimental tests of these complementarity relations applied to a particular class of one-qubit quantum states and also for random quantum states of one, two, and three qubits.
\end{abstract}


\maketitle

\section{Introduction}
\label{intro}
Wave-particle duality was one of the cornerstones in the development of Quantum Mechanics. This intriguing aspect is generally captured, in a qualitative way, by Bohr's complementarity principle \cite{Bohr}. It states that quantons \cite{Leblond} have characteristics that are equally real, but mutually exclusive.  Quoting Bohr \cite{Wit}: \textit{"... evidence  obtained  under different experimental conditions  cannot  be  comprehended  within  a single  picture,  but  must  be  regarded  as  complementary in  the  sense  that  only  the  totality  of  the  phenomena exhausts  the  possible  information  about  the  objects"}. For instance, in the Mach-Zehnder and double-slit interferometers,  the wave aspect is characterized by the visibility of interference fringes, meanwhile the particle nature is given by the which-way information of the path along the interferometer. In principle, the complete knowledge of the path destroys the interference pattern visibility, and vice-versa. 

However, in the quantitative scenario of the wave-particle duality explored by Wootters and Zurek \cite{Wootters}, where they investigated interferometers in which one obtains incomplete which-way information by introducing a path-detecting device, they showed that a partial interference pattern visibility can still be retained. Later, this work was extended by Englert, who derived a wave-particle duality relation \cite{Engle}. On the other hand, there is another way in which the wave-particle duality has been captured, without introducing path-detecting devices. Greenberger and Yasin \cite{Yasin}, considering a two-beam interferometer in which the intensities of the beams were not necessarily the same, defined a measure of path information called predictability. In this scenario, if the quantum system passing through the beam-splitter has different probabilities of getting reflected in the two paths, one has some path information about the quantum system. This line of reasoning resulted in a different kind of wave-particle duality relation
\begin{equation}
    P^2 + V^2 \le 1 \label{eq:cr1},
\end{equation}
where $P$ is the predictability and $V$ is the visibility of the interference pattern. Hence, by examining Eq. (\ref{eq:cr1}), one sees that even though an experiment can provide partial information about the wave and particle natures of a quantum system, the more information it gives about one aspect of the system, the less information the experiment can provide about the other. For instance, in Ref. \cite{Auccaise} the authors confirmed that it is possible to measure both aspects of the system with the same experimental apparatus, by using a molecular quantum information processor and employing Nuclear Magnetic Resonance techniques. 

In the last two decades, many authors have been taking steps towards the quantification of the wave-particle duality. D\"urr \cite{Durr}  and Englert et al. \cite{Englert} established minimal and reasonable conditions that any visibility and predictability measure should satisfy, and extended such measures for discrete $d$-dimensional quantons. Besides, with the development of the field of Quantum Information Science, it was suggested that quantum coherence \cite{Baumgratz} is a good generalization for the visibility measure \cite{Bera, Bagan, Tabish, Mishra}. Until now, many approaches were applied for quantifying the wave-particle properties of a quantum system \cite{Angelo, Coles, Hillery, Qureshi, Maziero}. It is worth mentioning that Baumgratz et al., in Ref. \cite{Baumgratz}, showed that the $l_1$-norm and the relative entropy of coherence are bona fide measures of coherence, meanwhile the Hilbert-Schmidt (or $l_2$-norm) coherence is not. However, as showed in Ref. \cite{Maziero}, all of these measures of quantum coherence, in addition to the Wigner-Yanase quantum coherence \cite{yu_Cwy}, are bone fide measures of visibility. Hence, one could expected that for each measure of coherence there exists a corresponding bona fide measure of predictability.

As pointed out by Qian et al. \cite{Qian}, complementarity relations like that in Eq. (\ref{eq:cr1}) do not really predict a balanced exchange between $P$ and $V$, once the inequality permits a decrease of $P$ and $V$ together, or an increase of both. It even allows the extreme case $P = V = 0$ to occur (neither wave nor particle) while, in an experimental setup, we still have a quanton on hands. Such a quanton can't be nothing. Thus, one can see that something must be missing from Eq. (\ref{eq:cr1}). As noticed by Jakob and Bergou \cite{Janos}, this lack of knowledge about the system is due to another intriguing quantum feature, entanglement \cite{Bruss, Horodecki}, or, more generally, to quantum correlations \cite{Marcos}. This means that information is being shared with another system, and this kind of quantum correlation can be seen as responsible for the loss of purity of each subsystem such that, for pure maximally entangled states, it is not possible to obtain information about the local properties of the subsystems. Therefore, to fully characterize a quanton, it is not enough to consider its wave-particle aspect; one has also to regard its correlations with other systems. 

Qian et al. also provided the first experimental confirmation of the complete complementarity relations using single photon states. Meanwhile, Ref. \cite{Schwaller} verified the link existing between entanglement and the amount of wave-particle duality with the superconducting qubits in the IBM Quantum Experience (IBMQE) \cite{ibmqe} for one particular bipartite quantum state. More recently, in Ref. \cite{Dittel}, the authors presented an architecture to investigate the wave-particle duality in $d$-path interferometers on a universal quantum computer involving as few as $2\log_2 (d)$ qubits, and developed a measurement scheme which allows for the efficient extraction of quantifiers of interference visibility and distinguishability. Lastly, as showed by two of us in Refs. \cite{Marcos, Basso}, if we consider the quanton as part of a multipartite pure quantum system, for each pair of coherence and predictability measures quantifying the local properties of a quanton, there is a corresponding quantum correlation measure that completes a given complementarity relation. More recently, we showed that for any complementarity relation of the type $P + V \le \text{constant}$, which saturates only for pure quantum system, with $P,V$ satisfying the criteria in \cite{Durr, Englert}, it follows that the corresponding quantum correlations are entanglement monotones, which can be extended for the mixed case through the convex roof method \cite{Leopoldo}.

The remainder of this article is organized as follows. In Sec. \ref{sec:cr}, we introduce the complementarity relations we verify experimentally in this article, together with the associated visibility, predictability, and quantum correlation measures. In Sec. \ref{sec:exp_setup} we describe some details of the experimental setup and related tools used for performing the experiments. In Sec. \ref{sec:res} we present the results of our experimental verification of the complementarity relations based on the properties of the density matrix. Sec. \ref{sec:res1} is dedicated to a particular class of one-qubit states while in Sec. \ref{sec:res2} we regard random quantum states of one, two, and three qubits. In Sec. \ref{sec:conc} we give our conclusions.

\section{Quantum coherence, predictability, and quantum correlation measures and the associated Complementarity relations}
\label{sec:cr}

In this section, we'll review some complementarity relations from the literature. We also report two new purity measures, in Eq. (\ref{eq:purity}). Let us consider a $n$-quanton pure quantum state described by $\ket{\Psi}_{A_1,...,A_n} \in \mathcal{H}_{1} \otimes ... \otimes \mathcal{H}_{n}$ with dimension $d = d_{A_1}d_{A_2}...d_{A_n}$. For each subsystem $A_m$, we define a local orthonormal basis in the Hilbert space $\mathcal{H}_m$ as $\{\ket{j}_{A_m}\}_{j = 0}^{d_{A_m} - 1}$, with $m = 1,\cdots,n$. The subsystem $A := A_1$ is represented by the reduced density operator $\rho_A = \Tr_{A_2,...,A_n} (\ket{\Psi}_{A_1,...,A_n} \bra{\Psi})$ \cite{Mark}. 
Starting from the purity of the global state $\ket{\Psi}_{A_1,...,A_n}$, it was shown in Refs. \cite{Marcos, Basso}
that the full characterization of the subsystem $A$ can be expressed by the following complete complementarity relations (CCRs):
\begin{align}
& P_{l_{1}}(\rho_A) + C_{l_{1}}(\rho_A) + W_{l_1}(\rho_A) = d_A - 1,  \label{eq:ccrl1} \\
    &  P_{hs}(\rho_A) + C_{wy}(\rho_A) + W_{wy}(\rho_A) = (d_A - 1)/d_A \label{eq:ccrwy},\\
    & P_{hs}(\rho_A) + C_{hs}(\rho_A) + S_l(\rho_A) = (d_A - 1)/d_A, \label{eq:ccrhs}\\
    & P_{vn}(\rho_A) + C_{re}(\rho_A) + S_{vn}(\rho_A) = \log_{2} d_A.  \label{eq:ccrre}  
\end{align}
The quantum coherence/visibility measures appearing in these CCRs are
\begin{align}
 & C_{l_{1}}(\rho_A) := \min_{\iota}||\rho_A-\iota||_{l_{1}} = \sum_{j \neq k} \abs{\rho^A_{jk}}, \\
    & C_{wy}(\rho_A)  := \sum_{j}I_{wy}(\rho_A,|j\rangle\langle j|) = \sum_{j\neq k}\abs{\bra{j}\sqrt{\rho_A}\ket{k}}^2, \\
    & C_{hs}(\rho_A) := \min_{\iota \in I}||\rho_A-\iota||_{hs}^{2} = \sum_{j \neq k} \abs{\rho^A_{jk}}^2,\label{eq:hsc} \\
    & C_{re}(\rho_A) := \min_{\iota \in I} S_{vn}(\rho_A||\iota) = S_{vn}(\rho_{A diag}) - S_{vn}(\rho_A), \label{eq:cre}
\end{align}
where $I$ is the set of all incoherent states, the Hilbert-Schmidt's norm of a matrix $M\in\mathbb{C}^{d \times d}$ is defined as $\norm{M}_{hs}:=  \sqrt{\sum_{j,k} \abs{M_{jk}}^2}$, whereas the $l_1$-norm is given by $\norm{M}_{l_1} := \sum_{j,k}\abs{M_{jk}}$, meanwhile $I_{wy}(\rho,\ketbra{j}) = -\frac{1}{2}\Tr ([\sqrt{\rho},\ketbra{j}]^2)$ is the Wigner-Yanase skew information, $S_{vn}(\rho_A||\iota) := \Tr(\rho_A \log_{2} \rho_A - \rho_A \log_{2} \iota)$ is the relative entropy, and $\rho_{Adiag}$ is the diagonal part of $\rho_{A}$. The predictability measures in the CCRs above are
\begin{align}
& P_{l_{1}}(\rho_A) := d_A - 1 - \sum_{j \neq k} \sqrt{\rho^A_{jj} \rho^A_{kk}}, \\
    & P_{hs}(\rho_A) := \sum_j (\rho^A_{jj})^2 - 1/d_A, \\
    & P_{vn}(\rho_A) :=  \log_{2} d_A + \sum_{j}\rho^A_{jj} \log_{2} \rho^A_{jj},
\end{align}
while the quantum correlation measures are
\begin{align}
& W_{l_1}(\rho_A) := \sum_{j \neq k}(\sqrt{\rho^A_{jj} \rho^A_{kk}} - \abs{\rho^A_{jk}}),\\
    & W_{wy}(\rho_A) := \sum_j( \bra{j}\sqrt{\rho_A}\ket{j}^2 - \bra{j}\rho_A\ket{j}^2),\\
    & S_l(\rho_A) := 1 - \Tr \rho_A^2 = 1 - \sum_{j,k} \abs{\rho^A_{jk}}^2,\\
    & S_{vn}(\rho_A) := - \Tr (\rho_A \log_{2} \rho_A),
\end{align}
where $\{|j\rangle\}_{j=0}^{d_{A}-1}\equiv\{|j\rangle_{A_{1}}\}_{j=0}^{d_{A}-1}$ and $\rho_{jk}^{A}=\langle j|\rho_{A}|k\rangle$.

It is worthwhile mentioning that the CCR in Eq. (\ref{eq:ccrhs}) is a natural generalization of the complementarity relation obtained by Jakob and Bergou for bipartite pure quantum systems \cite{Jakob, Bergou}. Besides, in Ref. \cite{Huber} the authors explored the purity-mixedness relation of a quanton to obtain a CCR equivalent to Eq. (\ref{eq:ccrhs}). We observe also that if the subsystem $A$ is not correlated with rest of the subsystems, then $A$ is pure and all the correlation measures vanish. In this case, $C_{wy}(\rho_A) = C_{hs}(\rho_A)$ and the CCRs in the Eqs. (\ref{eq:ccrwy}) and (\ref{eq:ccrhs}) become the same. In addition, we showed, in Ref. \cite{Basso}, that the quantum coherence measures used in this manuscript can be taken as quantum uncertainty measures, meanwhile the correlation measures can be taken as classical uncertainty measures. Therefore, the CCRs in Eqs. (\ref{eq:ccrl1}), (\ref{eq:ccrwy}), (\ref{eq:ccrhs}), and (\ref{eq:ccrre}) can be recast as complementarity relations between predictability and uncertainty, which implies that the predictability measures can be interpreted as measuring our capability of making a correct guess about the possible outcomes in the reference basis, i.e., if our total uncertainty about the possible outcomes decreases, our capability of making a correct guess has to increase. We can also see this by realizing that the expressions for $P_{vn}(\rho_A)$ and $P_{hs}(\rho_A)$ can be obtained from $P_{\tau}(\rho_A) = S^{max}_{\tau} -  S_{\tau}(\rho_{diag})$, $\tau = l, vn$, where $S_{\tau}(\rho_{diag})$ is measuring our total uncertainty about to possible outcomes. It worthwhile mentioning that for $d = 2$ we can write $P_{hs}(\rho) = \frac{1}{2}(\rho_{11} - \rho_{22})^2$, which is similar to predictability measure used in Refs. \cite{Durr, Englert}. Beyond that, it's possible to notice that $P = (f(\rho_{11}) - f(\rho_{22}))^2$ is also a bona-fide measure of predictability, with $f$ being any monotonic increasing function of the probabilities $\rho_{jj}, \  j= 1, 2$. Hence, for $f(x) =  \sqrt{x}$ the $l_{1}$-norm predictability is a generalization of two dimensional function $P = (\sqrt{\rho_{11}} - \sqrt{\rho_{22}})^2$. Lastly, we notice that all these visibility, predictability, and quantum correlation measures have the same physical significance, since they all meet the criteria established by the literature \cite{Durr, Englert}. Of course, this can change if an experiment or a physical situation that distinguishes them appears. If so, it will be necessary to modify or add some criteria to exclude some of the measures. 

Incomplete complementarity relations (ICRs) are obtained from Eqs. (\ref{eq:ccrl1}), (\ref{eq:ccrwy}), (\ref{eq:ccrhs}), and (\ref{eq:ccrre}) by ignoring the quantum correlations of the subsystem $A$ with the others subsystems:
\begin{align}
& P_{l_{1}}(\rho_A) + C_{l_{1}}(\rho_A) \le d_A - 1, \\
    &  P_{hs}(\rho_A) + C_{wy}(\rho_A)  \le (d_A - 1)/d_A \label{eq:crwy},\\
    & P_{hs}(\rho_A) + C_{hs}(\rho_A)  \le (d_A - 1)/d_A, \label{eq:crhs}\\
    & P_{vn}(\rho_A) + C_{re}(\rho_A) \le \log_{2} d_A,
\end{align}
since $S_{\tau}(\rho_A) \ge 0$ for $\tau = l, vn$ and $W_{\sigma}(\rho_A) \ge 0$ for $\sigma = l_1, wy$. These incomplete relations were also derived, in Ref. \cite{Maziero}, from the basic properties of the density matrix that describes the state of the subsystem. Moreover, ICRs describe the local aspects of a quanton, and are therefore closely linked to the purity of the system. For instance, the purity of the quantum system $A$ can be quantified by $\mathcal{P}_{hs}(\rho_A) = \Tr \rho^2_A$ \cite{Jaeger}, then it follows directly that $\mathcal{P}_{hs}(\rho_A) = P_{hs}(\rho_A) + C_{hs}(\rho_A) + 1/d_A$. In addition, as noticed in Ref. \cite{Gamel}, the von Neumann purity can be defined as $\mathcal{P}_{vn}(\rho_A) := \log_{2} d_A - S_{vn}(\rho_A)$, which implies $\mathcal{P}_{vn}(\rho_A) = P_{vn}(\rho_A) + C_{re}(\rho_A)$. Hence, it's suggestive to define 
\begin{equation}
\mathcal{P}_{\lambda}(\rho_A) := M_{\lambda}(d_A) - W_{\lambda}(\rho_A)
\label{eq:purity}
\end{equation}
as new measures of purity, where $M_{\lambda}(d_A) = d_A - 1, (d_A - 1)/d_A$ for $\lambda = l_1, wy$, respectively. 

We observe also that no system is completely isolated from its environment. This interaction between system and environment causes them to correlate, what leads to irreversible transference of information from the system to the environment. This process, called decoherence, results in a non-unitary dynamics for the system, whose most important effect is the disappearance of phase relationships between the subspaces of the system Hilbert space \cite{Zurek, Zurek1}. Therefore, the interaction between the quanton $A$ and the environment also introduces mixedness, as well in the rest of the system, which implies that the measures $S_l(\rho_A), S_{vn}(\rho_A), W_{l_1}(\rho_A), W_{wy}(\rho_A)$ can be seen, in general, as a measure of the mixedness of the subsystem $A$, since they do not distinguish the quantum correlations of the subsystem $A$ with rest of the subsystems from the undesirable correlations of the system with the environment.

\section{Experimental Setup}
\label{sec:exp_setup}
The IBM Quantum Experience (IBMQE) \cite{ibmqe} is a platform available to students and researchers from around the world which enables them to put quantum properties tests into practice by implementing quantum circuits on quantum chips. The quantum chips which are available have one, five, and fifteen qubits. In this work we use the quantum chips through the Qiskit platform, an Open-Source Quantum Development Kit, in which one can assemble the quantum circuit using Python programs. This facilitates changing the circuits parameters as we proceed with the experiments, and allows also for the use of the fifteen qubits quantum chip.

\begin{figure}
    \centering
    \caption{\label{config_chips} (Color online) Configuration of two quantum chips used in our experiments. In (a) we have the London chip, while in (b) we have the Yorktown chip configuration.}
    \includegraphics[width=0.48\textwidth]{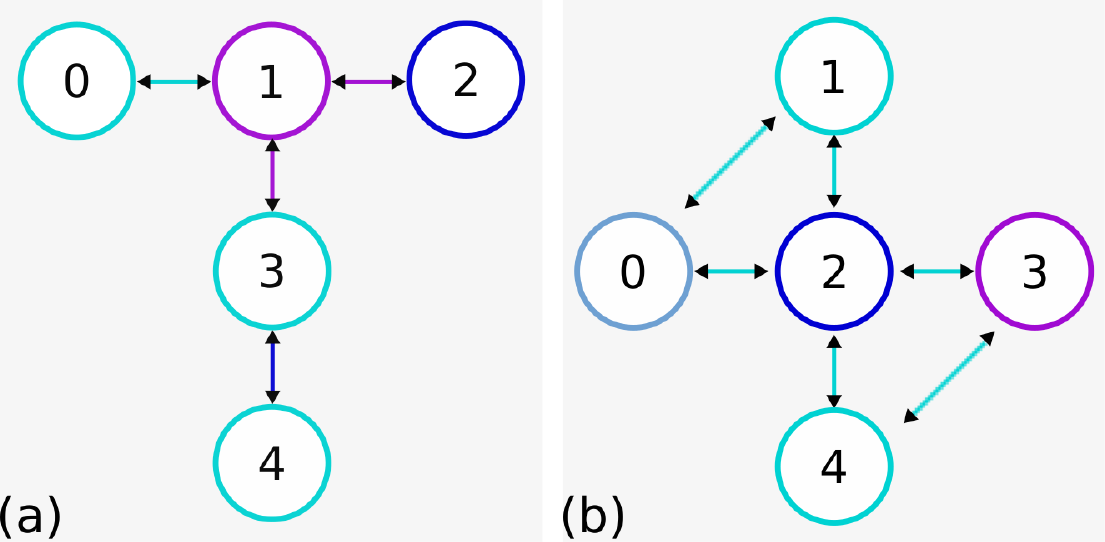}
\end{figure}

In our experiments, we used two different quantum chips, both with five qubits, the London and Yorktown chips. The configurations of these chips are shown in Fig. \ref{config_chips}. Their calibration parameters are presented in Tables \ref{yorktown_werner}, \ref{london_d2}, \ref{yorktwon_d4}, and \ref{yorktwon_d8}.
The temperature was $T = 0.0159 \text{ K}$ for both quantum chips in all experiments.

\begin{table}
\caption{\label{yorktown_werner} Calibration parameters for the Yorktown chip, when used for preparing the states of Sec. \ref{sec:res1}.}
\begin{tabular}{l c c}
\hline 
Yorktown parameters & Q0 & Q1 \tabularnewline
\hline 
\hline 
Frequency (GHz) & 5.28 & 5.25 \tabularnewline
T1 ($\mu$s) & 52.62 & 59.03 \tabularnewline
T2 ($\mu$s) & 22.88 & 26.51 \tabularnewline
Gate error ($10^{-3}$) & 1.81 & 1.02 \tabularnewline
Readout error ($10^{-2}$) & 6.20 & 2.40 \tabularnewline
Multiqubit gate error ($10^{-2}$) & $\stackrel{CX0\_1}{4.30}$ & $\stackrel{CX1\_0}{4.30}$ \tabularnewline
\hline
\end{tabular}
\end{table} 

\begin{table}
\caption{\label{london_d2} Calibration parameters for the London chip, when used for preparing the one qubit ($d=2$) random quantum states of Sec. \ref{sec:res2}.}
\begin{tabular}{l c c}
\hline 
 London parameters & Q0 & Q1 \tabularnewline
\hline 
\hline 
Frequency (GHz) & 5.25 & 5.05 \tabularnewline
T1 ($\mu$s) & 46.94 & 63.30 \tabularnewline
T2 ($\mu$s) & 76.55 & 50.48 \tabularnewline
Gate error ($10^{-4}$) & 5.03 & 3.41 \tabularnewline
Readout error ($10^{-2}$) & 2.50 & 3.50 \tabularnewline
Multiqubit gate error ($10^{-2}$) & $\stackrel{CX0\_1}{1.00}$ & $\stackrel{CX1\_0}{1.00}$ \tabularnewline
\hline
\end{tabular}
\end{table} 

\begin{table}
\caption{\label{yorktwon_d4} Calibration parameters for the Yorktown chip, when used for preparing the two qubit ($d=4$) random quantum states of Sec. \ref{sec:res2}.}
\begin{tabular}{l c c c}
\hline 
 Yorktown parameters & Q0 & Q1 & Q2 \tabularnewline
\hline 
\hline 
Frequency (GHz) & 5.29 & 5.24 & 5.03 \tabularnewline
T1 ($\mu$s) & 67.14 & 59.37 & 58.59 \tabularnewline
T2 ($\mu$s) & 81.44 & 62.05 & 59.08  \tabularnewline
Gate error ($10^{-4}$) & 5.56 & 10.11 & 5.83 \tabularnewline
Readout error ($10^{-2}$) & 1.50 & 1.55 & 2.35  \tabularnewline
Multiqubit gate error ($10^{-2}$) & $\stackrel{CX0\_1}{1.57}$ & $\stackrel{CX1\_0}{1.57}$ & $\stackrel{CX2\_0}{1.55}$ \tabularnewline
 & $\stackrel{CX0\_2}{1.55}$ &  $\stackrel{CX1\_2}{2.11}$ & $\stackrel{CX2\_1}{2.11}$
\tabularnewline
\hline
\end{tabular}
\end{table} 

\begin{table}
\caption{\label{yorktwon_d8} Average calibration parameters for the Yorktown chip, when used for preparing the three qubit ($d=8$) random quantum states of Sec. \ref{sec:res2}.}
\begin{tabular}{l c c c c}
\hline 
 Yorktown parameters & Q0 & Q1 & Q2 & Q3 \tabularnewline
\hline 
\hline 
Frequency (GHz) & 5.29 & 5.24 & 5.03 & 5.29\tabularnewline
T1 ($\mu$s) & 66.06 & 58.76 & 52.60 & 52.40\tabularnewline
T2 ($\mu$s) & 26.82 & 26.12 & 74.31 & 36.36 \tabularnewline
Gate error ($10^{-4}$) & 13.22 & 12.62 & 6.54 & 5.81\tabularnewline
Readout error ($10^{-2}$) & 5.52 & 2.87 & 2.38 & 1.33 \tabularnewline
Multiqubit gate error ($10^{-2}$) & $\stackrel{CX0\_1}{2.73}$ & $\stackrel{CX1\_0}{2.73}$ & $\stackrel{CX2\_0}{1.57}$ & $\stackrel{CX3\_2}{1.62}$ \tabularnewline
 & $\stackrel{CX0\_2}{1.57}$ &  $\stackrel{CX1\_2}{2.11}$ & $\stackrel{CX2\_1}{2.11}$ & $\stackrel{CX2\_3}{1.62}$
\tabularnewline
\hline
\end{tabular}
\end{table}

All qubits are always initialized in the state $|0\rangle$. After a quantum circuit is run, state tomography is performed using the function $\textit{state\underline\space tomography\underline\space circuits(qc,qr)}$, where $qc$ specifies the quantum circuit and $qr$ determines the qubits whose state is to be estimated. For the measurement of each observable mean value, needed for quantum state estimation, the circuit is run $8192$ times.

\section{Results}
\label{sec:res}
In this section, we report experimental results that verify the complementarity relations presented in Sec. \ref{sec:cr}. We start, in Sec. \ref{sec:res1}, considering a particular class of one-qubit states, what allows us to give a case study where visibility and predictability can diminish together. Afterwards, in Sec. \ref{sec:res2}, in order to report a more general verification of complementarity relations, we test them using random quantum states.

\subsection{A class of one-qubit states}
\label{sec:res1}

As mentioned before, complementarity relations represented by Eq. (\ref{eq:cr1}) do not really capture a balanced exchange between $P$ and $V$, because the inequality permits that $V$ decreases due to the interaction of the system with its environment, leading to the inevitable process of decoherence, while $P$ can remain unchanged or can even decrease together with the visibility of system. For instance, we can consider the following state for a quanton
\begin{equation}
    \rho_A = w \ket{\psi}_A \bra{\psi} + \frac{1 - w}{2}I_{2 \times 2}, \label{eq:wern}
\end{equation}
where $\ket{\psi}_A = \sqrt{x} \ket{0}_A + \sqrt{1 - x} \ket{1}_A$ with $x,w \in [0,1]$, and $I_{2 \times 2}$ is the identity operator. We can consider this state of system $A$ as the result of the interaction with its own environment modeled by the depolarizing channel \cite{nielsen}, which describes the situation wherein the interaction of the system with the surroundings mixes its state with the maximally entropic one. By inspecting Eq. (\ref{eq:wern}), we see that when $w \to 0$ the state of system $A$ approaches a maximally incoherent state, implying that $P \to 0$ and $V \to 0$ for any predictability and visibility measures. 

\begin{figure}
    \centering
    \caption{\label{circuito_werner} Quantum circuit we use to prepare the two-qubit state of Eq. (\ref{eq:psiAB}).}
    \includegraphics[width=0.4\textwidth]{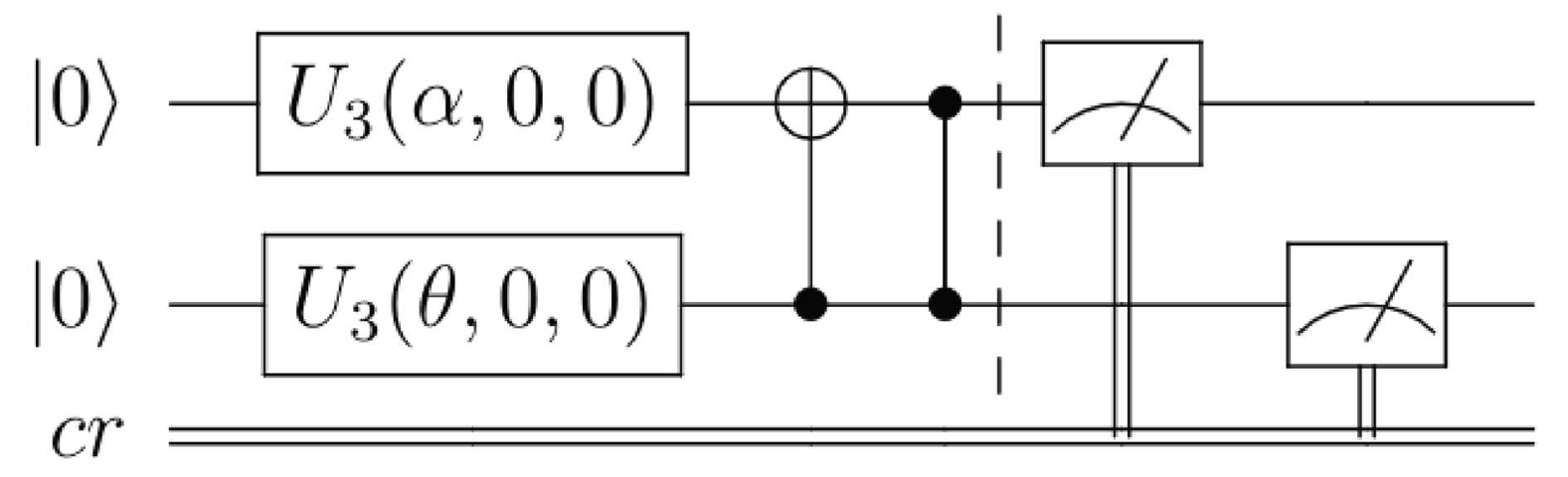}
\end{figure}

We can always purify $\rho_A$ and consider it as the result of entanglement with another system $B$, which may represent the degrees of freedom of the environment or of another auxiliary qubit. We notice that a possible purification for the state $\rho_{A}=\mathrm{Tr}_{B}|\psi\rangle_{AB}\langle\psi|$ is given as
\begin{align}
    \ket{\Psi}_{AB}& = (-\sqrt{1 - x}\ket{0}_A + \sqrt{x}\ket{1}_A) \otimes \sqrt{\frac{1 - w}{2}}\ket{0}_B \nonumber \\
    & + (\sqrt{x}\ket{0}_A + \sqrt{1 - x}\ket{1}_A) \otimes \sqrt{\frac{1 + w}{2}}\ket{1  }_B.
    \label{eq:psiAB}
\end{align}
We see that this pure state can be prepared experimentally using the following sequence of IBMQE unitary gates \cite{ibmqe}: 
\begin{align}
    \ket{\Psi}_{AB} & = C_{Z}(B \to A)C_{X}(B \to A)\nonumber \\ & \hspace{0.5cm}U^A_3(\alpha, 0,0) \otimes U^B_3(\theta, 0,0) \ket{0,0}_{A,B} \nonumber \\
    & = \Big( \cos \frac{\alpha}{2} \ket{0}_A + \sin \frac{\alpha}{2} \ket{1}_A \Big) \otimes \cos \frac{\theta}{2} \ket{0}_B \nonumber \\& + \Big( \sin \frac{\alpha}{2} \ket{0}_A - \cos \frac{\alpha}{2} \ket{1}_A \Big) \otimes \sin \frac{\theta}{2} \ket{1}_B,
\end{align}
where $U_{3}(\theta,\lambda,\phi)=\left[\begin{array}{cc} cos\frac{\theta}{2} & -e^{i\lambda}sen\frac{\theta}{2}\\ e^{i\phi}sen\frac{\theta}{2} & e^{i(\lambda+\phi)}cos\frac{\theta}{2} \end{array}\right]$ and $ C_{Z}(B \to A)=\Qcircuit @C=1em @R=1.0em {
& \ctrl{1} &  \qw  \\
& \ctrl{-1} &  \qw  
} = \left[\begin{array}{cccc} 
1 & 0 & 0 & 0   \\
0 & 1 & 0 & 0   \\
0 & 0 & 1 & 0   \\
0 & 0 & 0 & -1
\end{array}\right]$ and $C_{X}(B \to A)=\Qcircuit @C=1em @R=.7em {
& \targ & \qw \\ 
& \ctrl{-1} & \qw 
}= \left[ \begin{array}{cccc}
1 & 0 & 0 & 0\\ 
0 & 0 & 0 & 1\\
0 & 0 & 1 & 0\\ 
0 & 1 & 0 & 0 
\end{array} \right]$ are, respectively, the controlled-$Z$  and controlled-$X$ gates. Here
$B$ is the control qubit while $A$ is the target qubit \cite{Barenco}. A quantum circuit for preparing the state in Eq. (\ref{eq:wern}) is illustrated in Fig. \ref{circuito_werner}. The angles $\theta, \alpha$ and the parameters $w,x$ are related by $\alpha = 2 \arcsin (\sqrt{x})$ and $\theta = \arccos(-w)$, with $\alpha \in [0, \pi]$ and $\theta \in [\pi/2, \pi]$.

\begin{figure*}
    \centering
    \caption{\label{fig:3d} (Color online) (a) $C_{l1}$, (b) $P_{l1}$, (c) $W_{l1}$, and (d) $C_{l1} + P_{l1}$ along with $C_{l1} + P_{l1}+ W_{l1}$ of the state in Eq. (\ref{eq:wern}) prepared experimentally using the quantum circuit of Fig. \ref{circuito_werner}. The red lines represent the theoretical values, while the blue points are the experimental values. The error bars are the standard deviation for three repetitions of the experiment.}
    \includegraphics[width=1.0\textwidth]{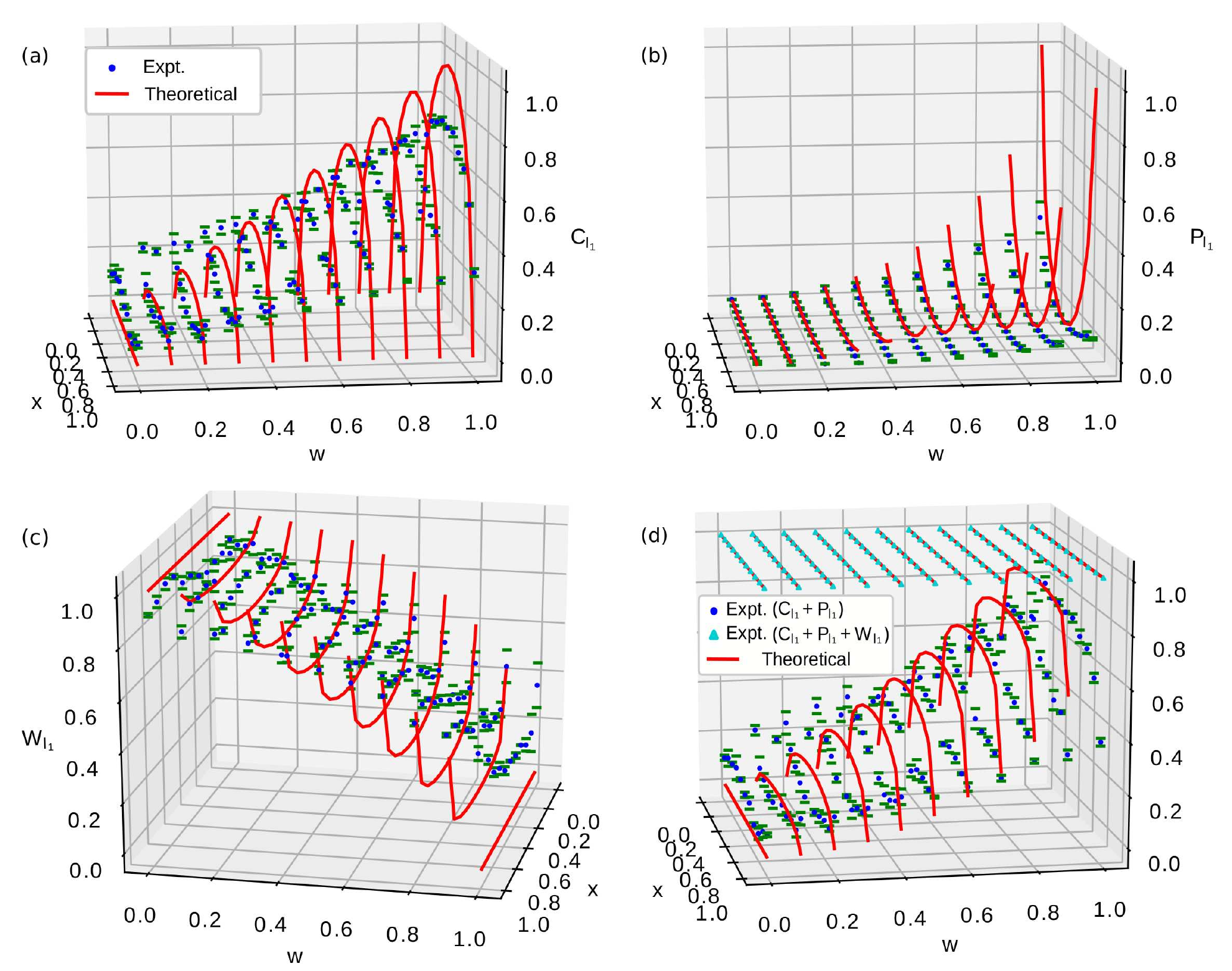}
\end{figure*}

Using this quantum circuit (implemented using the Yorktown chip with calibration parameters as shown in Table \ref{yorktown_werner}) to prepare $\rho_{A}$, in Fig. \ref{fig:3d} we report the experimental results of its quantum coherence, predictability, and quantum correlation measure, as well as for their sum. The experimental data points follow fairly well the theoretical predictions, with the predictability being the function most affected by the quantum computer imperfections and by its interaction with the environment. Most importantly, the complementarity relation $C_{l_{1}}(\rho_{A})+P_{l_{1}}(\rho_{A})\le d_{A}-1=1$ is satisfied for all experimental data points. Another interesting feature, which can be observed in Fig. \ref{fig:3d}, is that, in general, the experimental points for the quantum correlation measure exceeds the theoretical ones. This follows from the fact that the qubit $A$ is not correlated with just the auxiliary qubit $B$, but also with the environment, since, during the preparation of the state $\ket{\Psi}_{AB}$, the inescapable process of decoherence occurs. Therefore, we can see that the correlation measures do not distinguish the quantum correlations of the subsystem $A$ with rest of the subsystems from the inevitable correlations of the system with the environment.

\subsection{Random quantum states of one, two, and three qubits}
\label{sec:res2}

One can never provide a general verification of state dependent inequalities. However, the inclusion of randomness of the states used for the testing is arguably a good way forward advocating that such an inequality is satisfied generally, at least for the system dimensions regarded in those experimental tests. So, in this section we report experimental results verifying the complementarity relations of Sec. \ref{sec:cr} for random quantum states of one, two, and three qubits.

\begin{figure}
    \centering
    \caption{\label{fig:rcircuits} (Color online) Examples of (a) two-, (b) three-, and (c) four-qubit random quantum circuits that were executed on the IBM quantum computer chips.}
    \includegraphics[width=0.48\textwidth]{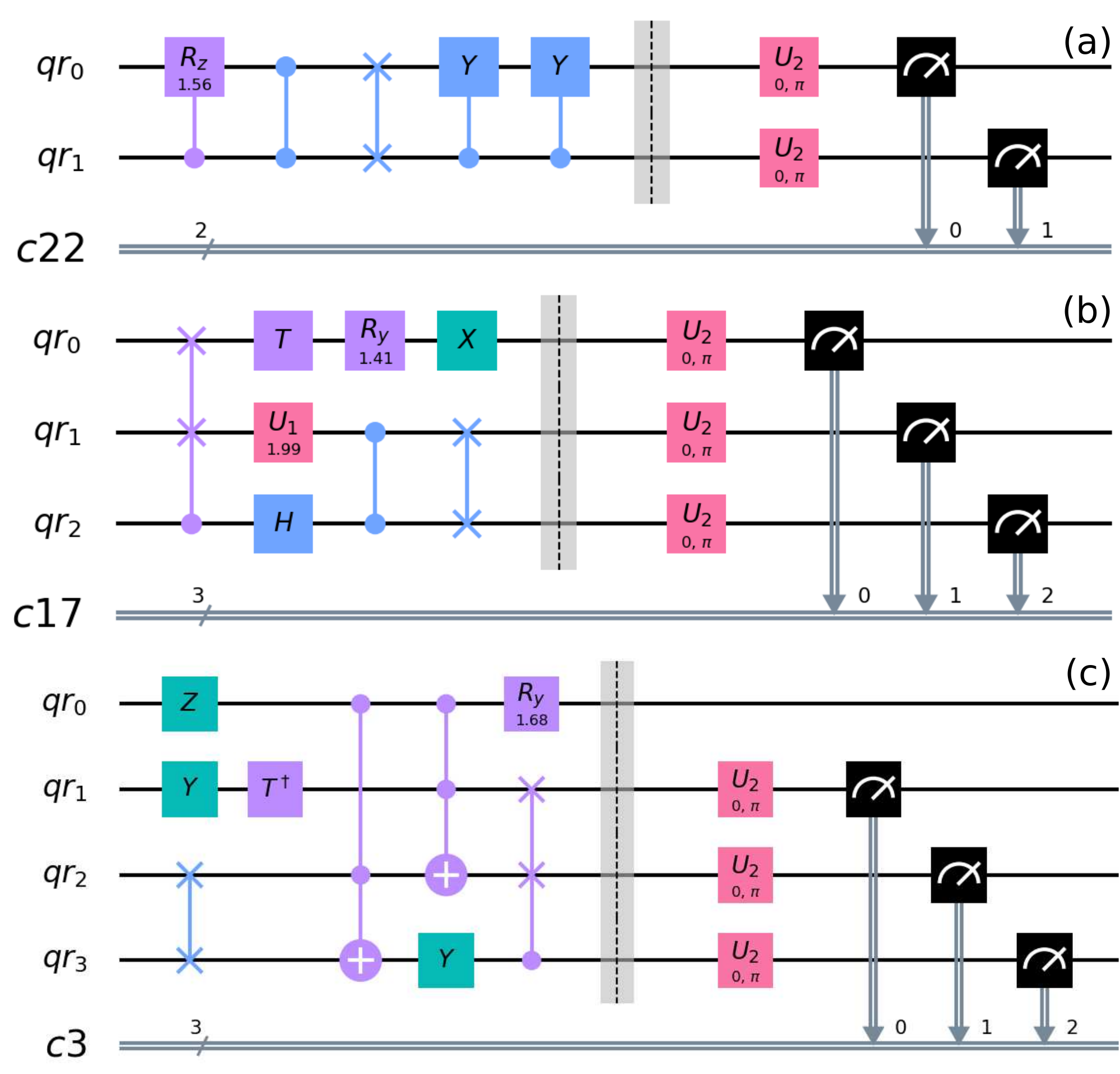}
\end{figure}

We use IBM's quantum computers aiming to prepare a pure state of $n+1$ qubits, $|\psi_{n+1}\rangle$. However, due to the quantum computer imperfections and its interaction with the surrounding environment, the state actually prepared will be a mixed state $\rho_{n+1}$.
From these density matrices, we obtain, via partial trace, the mixed states of $n$ qubits, $\rho_{n}=\mathrm{Tr}_{p}\rho_{n+1}$, that we use for experimentally verifying complementarity relations. After preparing the qubits in the state $|0\rangle$, a certain number of random quantum gates is applied to them. This is done using the Qiskit function $\textit{random\underline\space circuit(x,y)},$ where $x$ is the number of qubits to be included in the circuit and $y$ is the number of random quantum gates to be applied to them. We observe that the quantum gates are sampled randomly from the IBMQE set of quantum gates \cite{ibmqe}, with all gates having the same probability of being selected. Some examples of these circuits are shown in Fig. \ref{fig:rcircuits}. The quantum gates used in the random quantum circuits are listed below. Besides those already mentioned in Sec. \ref{sec:res1}, we have the Pauli gates:
$\Qcircuit @C=1em @R=1.0em {
& \gate{X} &\qw 
}= \left[\begin{array}{cc} 0 & 1 \\
1 & 0 \end{array}\right]$,
$\Qcircuit @C=1em @R=1.0em {
&  \gate{Y} & \qw 
}=\left[\begin{array}{cc} 0 & -i \\ 
i & 0 \end{array}\right]$, and
$\Qcircuit @C=1em @R=1.0em {
&  \gate{Z} &\qw 
} = \left[\begin{array}{cc} 1 & 0 \\
0 & -1 \end{array}\right],$ the identity gate
$\Qcircuit @C=1em @R=1.0em {
&  \gate{I} &\qw 
} = \left[\begin{array}{cc} 1 & 0 \\
0 & 1 \end{array}\right],$
the Hadamard gate
$\Qcircuit @C=1em @R=1.0em {
& \gate{H} &\qw 
} =  \frac{1}{\sqrt{2}}\left[\begin{array}{cc} 1 & 1\\ 1 & -1 \end{array}\right] $, the rotation gates 
$\Qcircuit @C=1em @R=1.0em {
& \gate{R_{X}(\theta)} &\qw 
} = \left[\begin{array}{cc} 
\cos(\theta/2) & -i\sin(\theta/2) \\ 
-i\sin(\theta/2) & \cos(\theta/2) 
\end{array}\right] $, 
$\Qcircuit @C=1em @R=1.0em {
& \gate{R_{Y}(\theta)} &\qw 
} = \left[\begin{array}{cc} 
\cos(\theta/2) & -\sin(\theta/2) \\
\sin(\theta/2) & \cos(\theta/2) 
\end{array}\right] $,
$\Qcircuit @C=1em @R=1.0em {
& \gate{R_{Z}(\phi)} &\qw 
} = \left[\begin{array}{cc} 
e^{-i(\phi/2)} & 0\\ 
0 & e^{i(\phi/2)}
\end{array}\right],$ and the phase gates
$\Qcircuit @C=1em @R=1.0em {
& \gate{S} &\qw 
} = \left[\begin{array}{cc} 1 & 0 \\
0 & i \end{array}\right]$,
$\Qcircuit @C=1em @R=1.0em {
& \gate{S^{\dagger}} &\qw 
} = \left[\begin{array}{cc} 1 & 0 \\ 
0 & -i \end{array}\right]$,
$\Qcircuit @C=1em @R=1.0em {
& \gate{T} &\qw 
} = \left[\begin{array}{cc} 1 & 0 \\ 
0 & \frac{1+i}{\sqrt{2}} \end{array}\right]$,
$\Qcircuit @C=1em @R=1.0em {
&  \gate{T^{\dagger}} &\qw 
}= \left[\begin{array}{cc} 1 & 0 \\
0 & \frac{1-i}{\sqrt{2}} \end{array}\right].$
The gates $\Qcircuit @C=1em @R=.7em { & \gate{U_{1}(\lambda)} & \qw } = \left[\begin{array}{cc} 1 & 0\\ 0 & e^{i\lambda} \end{array}\right]$and $\Qcircuit @C=1em @R=.7em { & \gate{U_{2}(\lambda,\phi)} & \qw } = \left[\begin{array}{cc} 1 & -e^{i\lambda}\\ e^{i\phi} & e^{i(\lambda+\phi)} \end{array}\right]$ are similar to $U_{3}$, but with fewer parameters.
The SWAP gate,
$\Qcircuit @C=1em @R=1.0em { & \qswap &  \qw \\
& \qswap \qwx & \qw
} = \left[\begin{array}{cccc} 
1 & 0 & 0 & 0  \\
0 & 0 & 1 & 0   \\
0 & 1 & 0 & 0   \\
0 & 0 & 0 & 1   
\end{array}\right],$  exchanges the states of two qubits. The Toffoli gate uses two control qubits and one target qubit:
$\Qcircuit @C=1em @R=1.0em {
& \ctrl{1} &  \qw  \\
& \ctrl{1} &  \qw  \\
& \targ    &  \qw
} = \left[\begin{array}{cccccccc} 
 1 & 0 & 0 & 0 & 0 & 0 & 0 & 0\\ 
 0 & 1 & 0 & 0 & 0 & 0 & 0 & 0\\
 0 & 0 & 1 & 0 & 0 & 0 & 0 & 0\\
 0 & 0 & 0 & 0 & 0 & 0 & 0 & 1\\
 0 & 0 & 0 & 0 & 1 & 0 & 0 & 0\\
 0 & 0 & 0 & 0 & 0 & 1 & 0 & 0\\
 0 & 0 & 0 & 0 & 0 & 0 & 1 & 0\\
 0 & 0 & 0 & 1 & 0 & 0 & 0 & 0\\
 \end{array}\right].$
The gate 
$\Qcircuit @C=1em @R=1.0em {
& \gate{\bullet} &\qw 
}$ represents the controller of any port, which acts on a target qubit depending on the state of the control qubit. The barrier $\Qcircuit @C=1em @R=1.0em { & \qw  \barrier[-0.95em]{1} &\qw \\ & \qw & \qw  \\ }$ has the function of optimizing compilation and improving the visualization of the circuit. Finally, $\Qcircuit @C=1em @R=.7em { & \meter & \qw }$ performs one-qubit measurements on the standard basis.

\begin{figure}[]
    \centering
    \caption{\label{cp_dimension} (Color online) (a) $C_{l1}$, (b) $P_{l1}$, (c) $W_{l1}$, and $C_{l1}+P_{l1}$ and $C_{l1}+P_{l1}+W_{l1}$ as a function of the system dimension $2^{n}$, where $n=1,2,3$ is the number of qubits. The values marked with the red "\textbf{x}" symbols are the theoretical values while the values marked with the blue "$\bullet$" symbols are the experimental ones. The horizontal bar, for each dimension, represents $C_{l1} + P_{l1}+W_{l_{1}}=d_{A}-1,$ the upper bound  for the sum $C_{l1} + P_{l1}$.}
    \includegraphics[width=0.43\textwidth]{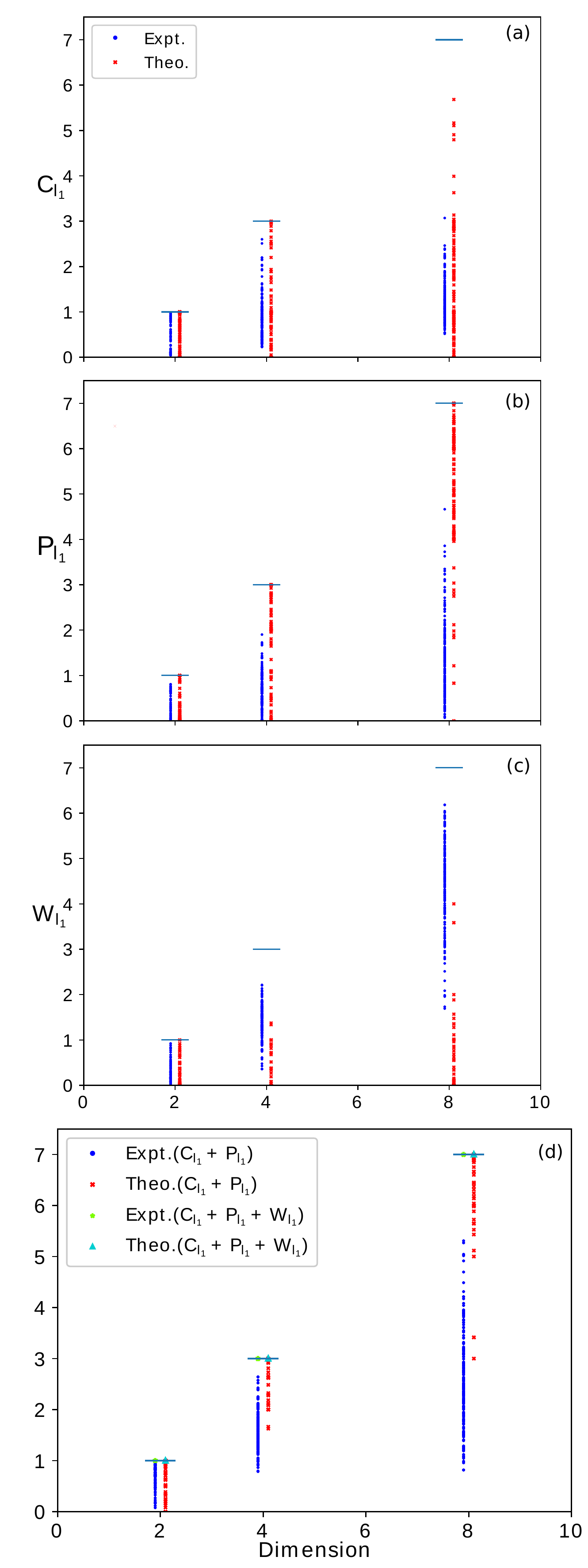}
\end{figure}

Using random quantum states of one ($d=2$), two ($d=4$), and three ($d=8$) qubits, we present in Fig. \ref{cp_dimension} the experimental verification of the complementarity relations listed in Sec. \ref{sec:cr}. For $d=2$, we used the London ship with the calibration parameters of Table \ref{london_d2}. We prepared $100$ random quantum states using a circuit composed by five random quantum gates (this circuit is applied to $n+1=2$ qubits). For dimension $d=4$ we used the Yorktown chip with the calibration parameters shown in Table \ref{yorktwon_d4}. In this case we prepared $150$ random quantum states via circuits with four random quantum gates. At last, for $d=8$ we applied the Yorktown chip with average calibration parameters as shown in Table \ref{yorktwon_d8}. The average was taken over three different calibrations. Here $200$ random quantum states were prepared using random quantum circuits composed by four quantum gates.

We see, in Fig. \ref{cp_dimension}, that the complementarity relation $C_{l_{1}}+P_{l_{1}}\le d_{A}-1$ is also verified by our experimental results for random quantum states. As in the previous section, in this case the interaction with the environment, and the consequent creation of quantum correlation between system and environment, also diminishes the sum of coherence and predictability. This explains the general ``shift'' to lower values of the experimental values of $C_{l_{1}}+P_{l_{1}}$ when compared with the theoretical predictions.

Finally, we observe that the results for the complementarity relations associated to Eqs. (\ref{eq:ccrwy}), (\ref{eq:ccrhs}), and (\ref{eq:ccrre}) are qualitatively similar to the ones shown in Figs. \ref{fig:3d} and \ref{cp_dimension}. So, all complementarity relations presented in Sec. \ref{sec:cr} are validated by our experimental results.

\section{Final remarks}
\label{sec:conc}
Bohr's complementarity principle accompanied Quantum Mechanics since its early inception. Even though for a long time this principle was thought to be a result to be added to the quantum formalism, recently it was shown that the first can be derived from the later. In this article, besides introducing two new purity measures, we experimentally verified complementarity relations that are based on the fundamental properties of the quantum density matrix: positivity and unit trace. Besides using a class of one-qubit states to highlight the importance of considering complete complementarity relations, we reported the experimental verification of complementarity relation for random one-, two-, and three-qubit states. In both scenarios, we showed that the interaction with the environment, and the consequent creation of quantum correlation between system and environment, diminishes the sum of coherence and predictability, what explains the generally lower experimental data values when compared with the theoretical predictions. On the other hand, as expected, the experimental values of the correlation measures are generally bigger when compared with the theoretical ones. Nevertheless, the complete and incomplete complementarity relations are always satisfied, once the lower values of coherence and predictability are balanced by the bigger values of the correlation measure.

\begin{acknowledgments}
This work was supported by the Coordena\c{c}\~ao de Aperfei\c{c}oamento de Pessoal de N\'ivel Superior (CAPES), processes 88882.427924/2019-01 and 88882.427897/2019-01, and by the Instituto Nacional de Ci\^encia e Tecnologia de Informa\c{c}\~ao Qu\^antica (INCT-IQ), process 465469/2014-0.
\end{acknowledgments}


\end{document}